\documentclass[final]{aipproc}

\layoutstyle{6x9}

\newcommand{\p}{\prime}
\newcommand{\sumi}{\sum_{i=0,\,\mathrm{even}}^{n-1}}

\begin{document}
\title{Understanding Parton Distributions from\\Lattice QCD}
\author{Dru B. Renner}{
address={Department of Physics,
University of Arizona,
1118 E 4th Street,
Tucson, AZ 85721, USA}
}
\begin{abstract}
I examine the past lattice QCD calculations of three representative 
observables, the transverse quark distribution, momentum fraction, and 
axial charge, and emphasize the prospects for not only quantitative 
comparison with experiment but also qualitative understanding of QCD.
\end{abstract}
\maketitle
\section{Introduction}
Lattice QCD calculations provide the opportunity for both quantitative comparison
with experimental measurements and for advancing our qualitative understanding 
of QCD.  I examine several observables which exemplify this range of opportunities.
The lattice calculation of the moments of parton distributions
%of the momentum fraction and axial charge and remaining moments of parton distributions
is an essential step in achieving quantitative agreement with experimental results. 
Additionally our understanding of QCD can be further expanded by calculating
%novel observables such as the moments of generalized parton distributions
the remaining moments of the generalized parton distributions which determine the
three dimensional distribution of the transverse position and longitudinal 
momentum of quarks and gluons within the nucleon.
\paragraph{Generalized Form Factors}
The generalized form factors provide an alternative but equivalent language to
the generalized parton distributions.  They encode the ordinary form 
factors and parton distributions as well as the nucleon spin decomposition~\cite{Ji:1996ek}
and transverse quark distributions~\cite{Burkardt:2000za} and hence provide a unifying language 
to describe calculations of nucleon structure.  Each tower of twist two operators has a 
corresponding set of generalized form factors.
As an example the unpolarized operators
$O_q^{\mu_1 \cdots \mu_n} = \overline{q} iD^{(\mu_1} \cdots iD^{\mu_{n-1}} \gamma^{\mu_n)} q$
define the generalized form factors $A^q_{ni}$, $B^q_{ni}$, and $C^q_{n}$ via
\begin{equation}
\label{gffs}
\langle P^\p | O_q^{\mu_1 \ldots \mu_n} | P \rangle = 
\overline{U}(P^\p) [ \sumi \left( A^q_{ni}(t) K^A_{ni} + B^q_{ni}(t) K^B_{ni} \right)
                              + \delta^n_\mathrm{even} C^q_n(t)    K^C_{n}\,\, ] U(P)
\end{equation}
where K are known functions of $P$ and $P^\p$. A complete set of results can
be found in \cite{Diehl:2003ny}.
%Most importantly the generalized
%form factors are defined in terms of matrix elements of local operators which
%can be calculated in Euclidean space on the lattice.
%
\paragraph{Lattice Calculations}
There have been several full QCD calculations of nucleon structure to 
date.  Results from these and other calculations [4-14]
%~\cite{:2003is,Gockeler:2004vx,Renner:2005sm,Gockeler:2002ek,Detmold:2001jb,Ohta:2004mg,Sasaki:2003jh,Khan:2004vw,Renner:2004ck}
were shown at the conference.
Each calculation uses different actions as well as differing lattice spacings and volumes.
However the dominate systematic error in lattice calculations of 
nucleon observables is still due to the chiral extrapolation. 
% Calculations
%using improved staggered actions provide the best tool currently available
%to reach light quark masses, however, dynamical domain wall calculations 
%may ultimately be competetive.
%The details of the calculations discussed here can be found in the above references.
%
\section{Transverse Quark Distributions}
%The last few years have seen the first lattice calculations of the
%nucleon's generalized form factors which represent an opportunity 
%to enhance both our quantitative and qualitative understanding by 
%examining the transverse structure of the nucleon.
\begin{ltxfigure}[t]
\begin{minipage}{16pc}
\includegraphics[width=16pc]{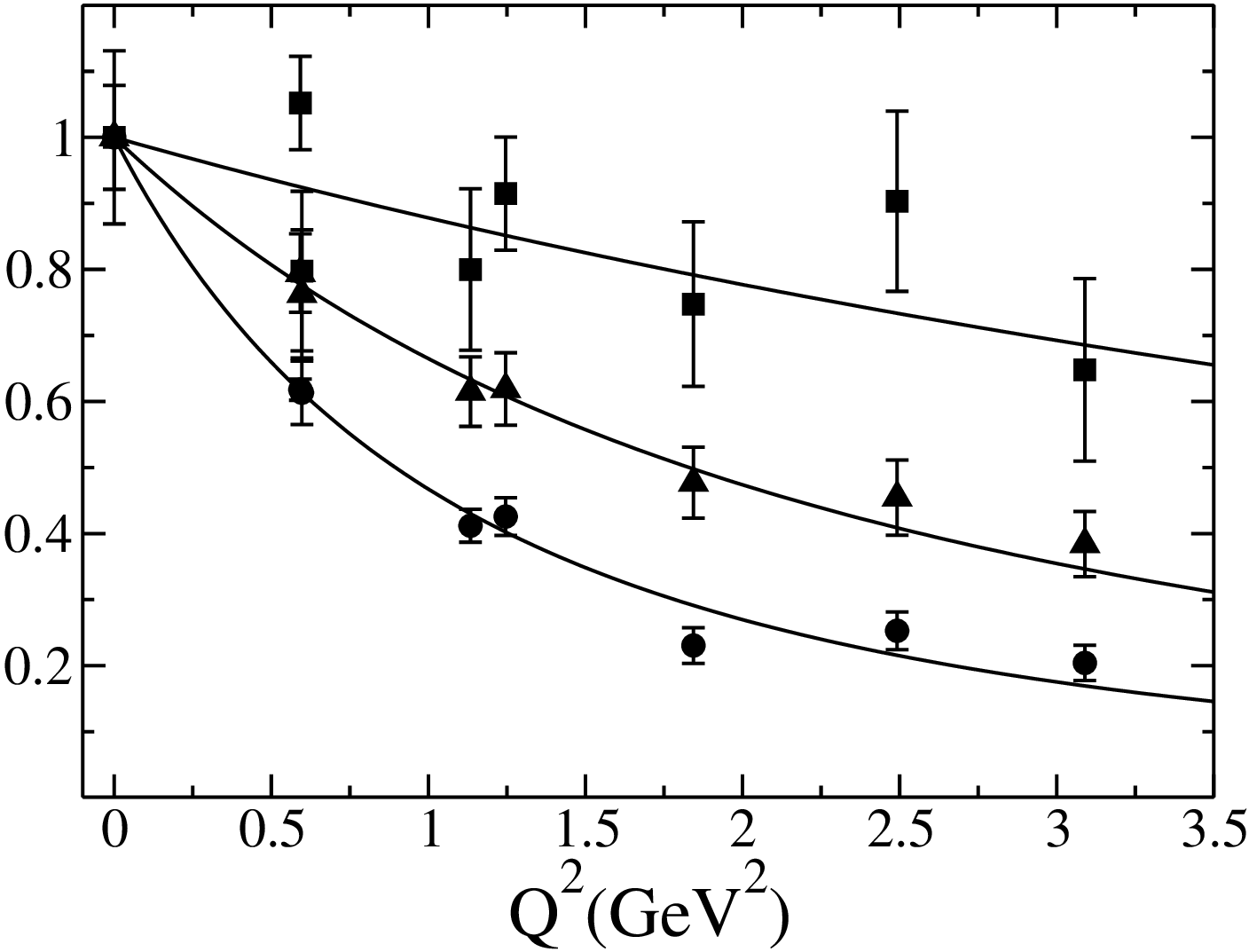}
\caption{\label{AAA}squares are $A^{u-d}_{30}$, triangles are $A^{u-d}_{20}$, circles are $A^{u-d}_{10}$ \cite{:2003is}}
\end{minipage}\hspace{2pc}
\begin{minipage}{16pc}
\includegraphics[width=16pc]{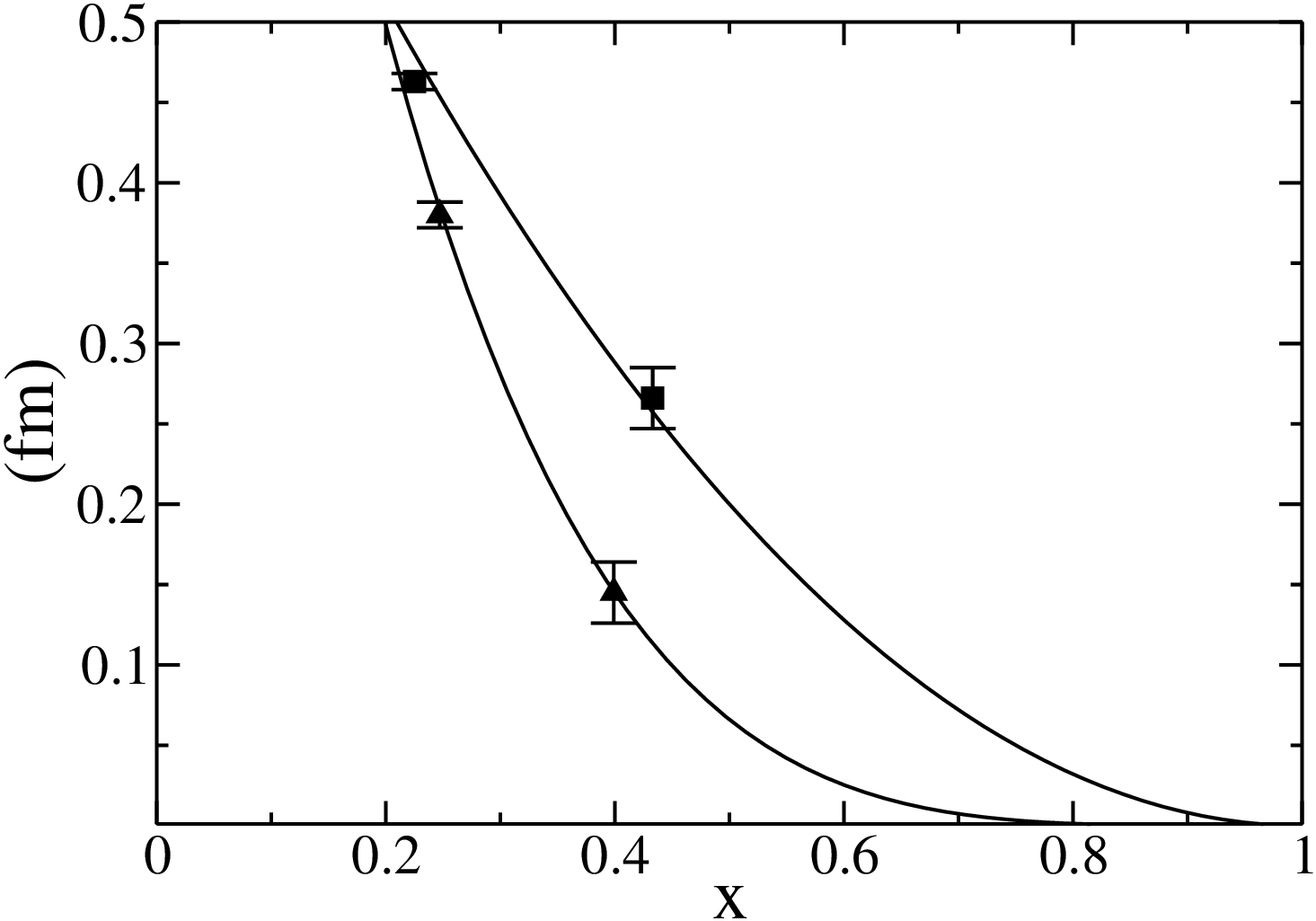}
\caption{\label{rms}squares are $\langle b^2_\perp \rangle^{u+d}_{(n)}$, triangles are $\langle b^2_\perp \rangle^{u-d}_{(n)}$ for $n=1,3$ \cite{:2003is,Renner:2005sm}}
\end{minipage}
\end{ltxfigure}
The transverse 
quark distribution $q(x,\vec{b}_\perp)$ gives the probability to
find a quark of flavor $q$ carrying a fraction $x$ of the nucleon's
longitudinal momentum at a displacement $\vec{b}_\perp$ from the 
center of the nucleon.  The moments of the transverse quark distribution
are given by 
\begin{equation}
\label{mom}
q_n(\vec{b}_\perp) = \int_{-1}^{1}\!dx\,\, x^{n-1}\, q(x,\vec{b}_\perp) = \int \frac{d^2\Delta_\perp}{(2\pi)^2} e^{-i\vec{b}_\perp\cdot\vec{\Delta}_\perp} A^q_{n0}(-\vec{\Delta}_\perp^2).
\end{equation}
Current calculations are restricted to the lowest three moments, however
the following
%paragraphs
shows several ways to examine 
%in the following we show several ways to examine 
the transverse structure
%of the nucleon 
using just the low moments.
\paragraph{$Q^2$ Dependence}
%The generalized form factors are calculated directly from lattice QCD as
%functions of $Q^2$.
The slope of the generalized form factors in the forward limit is of particular
interest because it determines the rms radius of the $n{}^\mathrm{th}$ moment of
$q(x,\vec{b}_\perp)$,
\begin{equation}
\label{rmseq}
\left< b_{\perp}^2 \right>_{n} =
\frac{ \int\!d^2b_{\perp}\, b_{\perp}^2\, \int_{-1}^{1}\! dx\, x^{n-1}q(x,\vec{b}_{\perp}) }
{ \int\!d^2b_{\perp}\, \int_{-1}^{1}\! dx\, x^{n-1}q(x,\vec{b}_{\perp}) }
=\frac{-4}{A^q_{n0}(0)}\frac{dA^q_{n0}(0)}{dQ^2}.
\end{equation}
The moments in Eq.~\ref{rmseq} are dominated by $x$ near $1$ for large $n$.
In such a limit the quark carries all the longitudinal momentum
and is kinematically constrained to reside at the center of the
nucleon~\cite{:2003is}.
Thus higher moments determine the transverse size of larger $x$ quarks which 
are distributed more narrowly in $b_\perp$.  Consequently 
the qualitative expectation, illustrated in Fig.~\ref{AAA}, is that the
slopes
%of the generalized form factors
should decrease as $n$ increases for large
enough $n$.
%This expectation is illustrated in Fig.~\ref{AAA}.
%and is observed in
%the calculations by both LHPC and QCDSF.
That
an expectation for large $n$ is so clearly seen for the lowest three moments
%is a bit surprising but
demonstrates that the transverse distribution of quarks 
within the nucleon depends strongly on the momentum fraction at which the quarks
are probed.
%This conclustion is strengthened in the following paragraphs.
%
\paragraph{$x$ Dependence}
The transverse rms radius of the nucleon at a fixed $x$ is
%momentum fraction $x$ is
%
\begin{displaymath}
\left< b_{\perp}^2 \right>_x =
\frac{ \int\!d^2b_{\perp}\, b_{\perp}^2\, q(x,\vec{b}_{\perp}) }
{ \int\!d^2b_{\perp}\, q(x,\vec{b}_{\perp}) },
\end{displaymath}
whereas lattice calculations determine the transverse radius
at a fixed moment $n$ as shown in Eq.~\ref{rmseq}.
%In an attempt
%to create an intuitive understanding of the nucleon's transverse 
%structure
To understand the meaning of the transverse radius at a fixed moment 
we can think of $\left<b^2_\perp\right>_n$ as a 
coarse grained transverse size of the nucleon corresponding
to a region centered on the average $x$ of the $n^\mathrm{th}$ moment,
\begin{equation}
\label{xave}
\left<x\right>_n =
\frac{ \int\!d^2b_{\perp}\, \int_{-1}^{1}\! dx\, \left|x\right| x^{n-1}q(x,\vec{b}_{\perp}) }
{ \int\!d^2b_{\perp}\, \int_{-1}^{1}\! dx\, x^{n-1}q(x,\vec{b}_{\perp}) }
= \frac{\left<x^n\right>+2(-1)^n\int\!d^2b_{\perp}\int_{0}^{1}\!dx\,x^n\overline{q}(x)}{\left<x^{n-1}\right>}.
%%\approx \frac{<x^n>}{<x^{n-1}>}.
\end{equation}
Lattice QCD is not currently 
capable of calculating the anti-quark contribution in $\left<x\right>_n$,
however the phenomenologically determined parton distributions indicate it is small
enough that it does not affect the following qualitative conclusions.  Fig.~\ref{rms}
shows the transverse radius versus corresponding momentum fraction for the
lowest moments illustrating, as above, that the 
transverse size of the nucleon depends significantly on 
the longitudinal momentum of its constituents.
\paragraph{$b_\perp$ Dependence}
By assuming a dipole ansatz for the generalized form factors the $b_\perp$ 
dependence of each moment can be determined from Eq.~\ref{mom}.  The details
are given in~\cite{Renner:2005sm}, and the results are shown in Fig.~\ref{qqq}.
Of particular importance are the lowest two moments which determine the
transverse distribution of quarks ($n=1$) and of longitudinal momentum ($n=2$) within the
nucleon.
\begin{ltxfigure}[t]
\begin{minipage}{16pc}
\includegraphics[width=16pc]{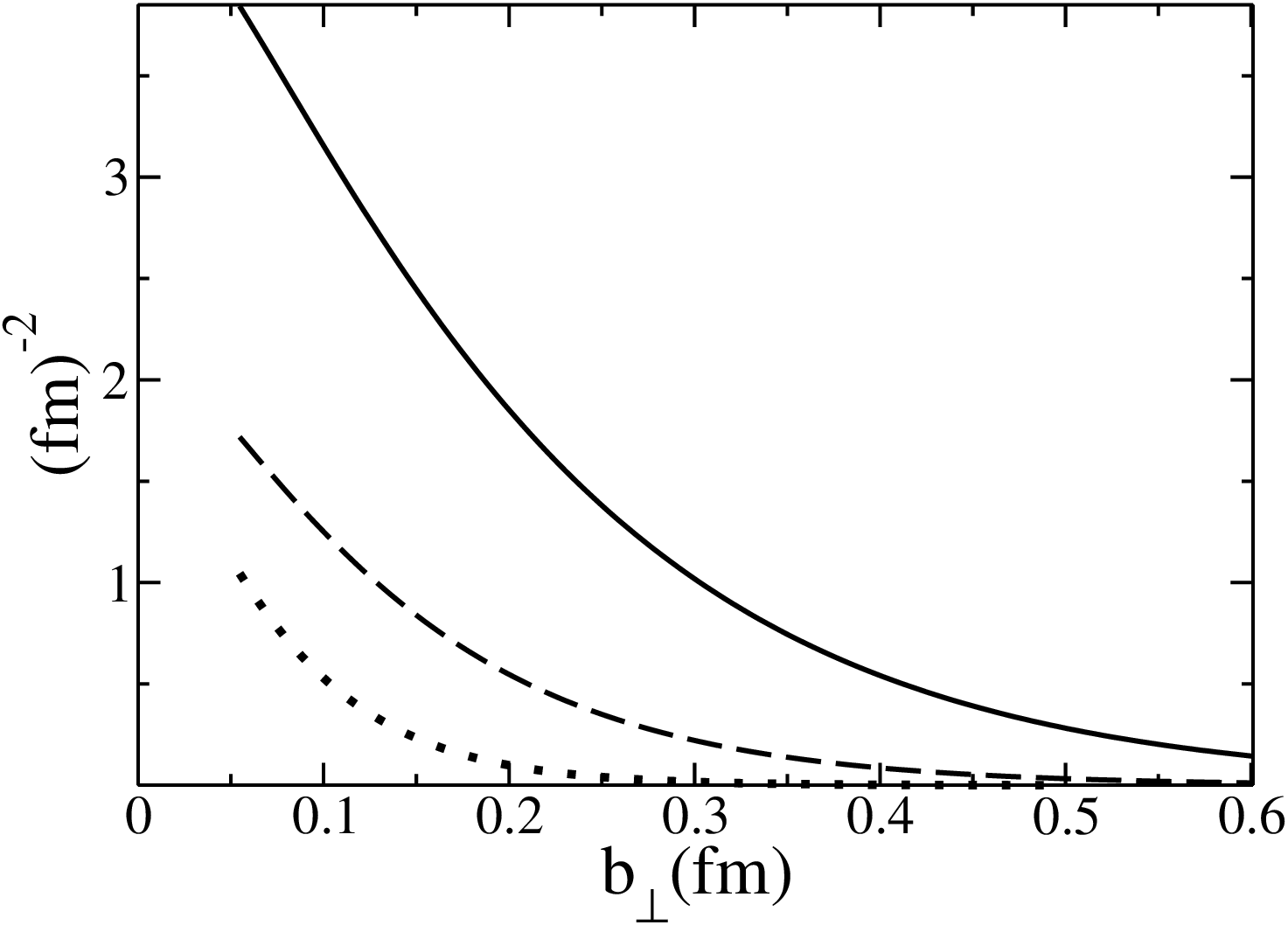}
\caption{\label{qqq}solid, dashed, dotted are $q_n(\vec{b}_\perp)$ for $n=1,2,3$, $q=u-d$ \cite{Renner:2005sm}}
\end{minipage}\hspace{2pc}
\begin{minipage}{16pc}
\includegraphics[width=16pc]{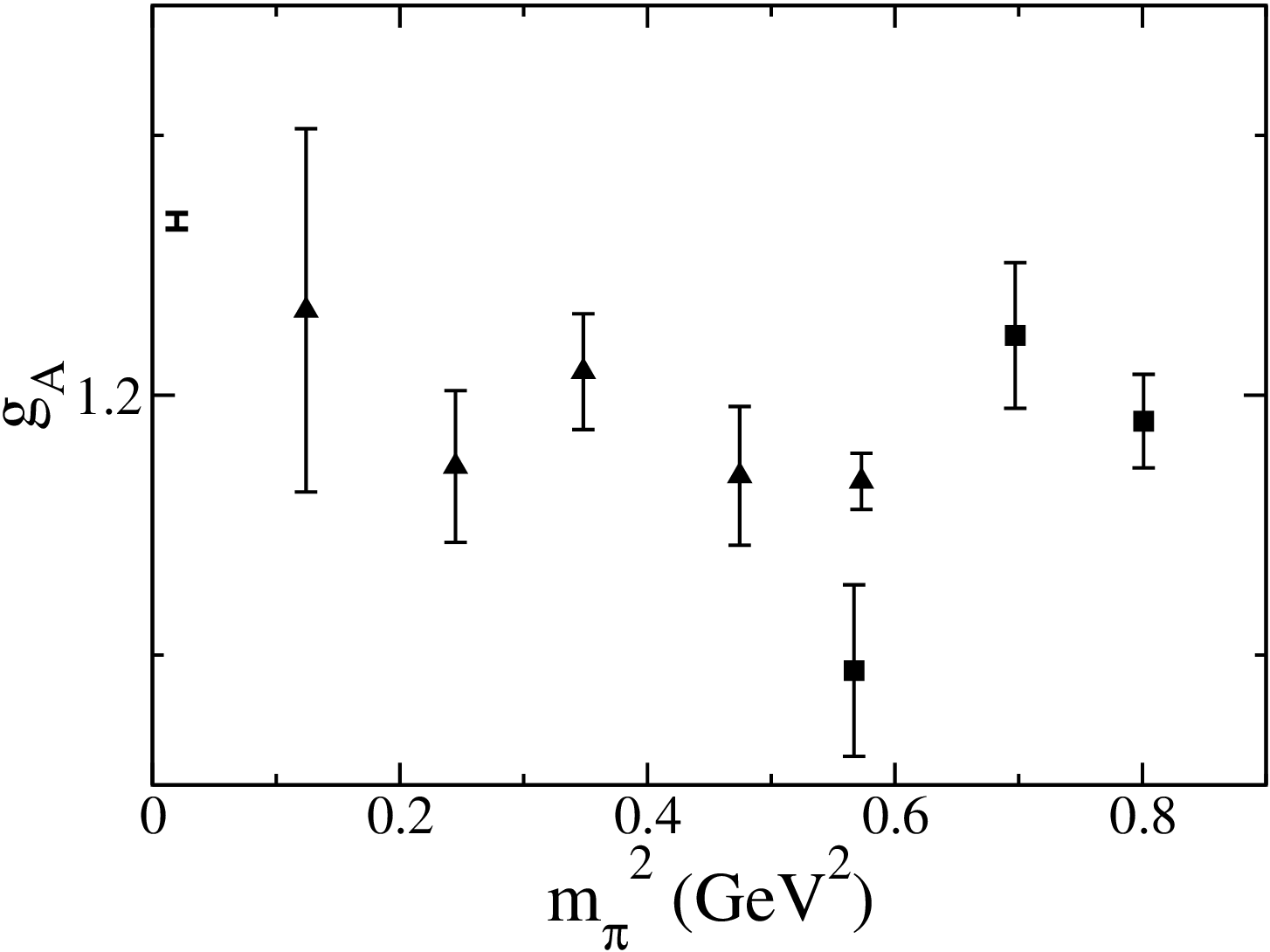}
\caption{\label{gA}$g_A$, triangles~\cite{Renner:2004ck}, squares~\cite{Dolgov:2002zm}}
\end{minipage}
\end{ltxfigure} 
\section{Moments of Parton Distributions}
Moments of parton distributions provide an opportunity for quantitative
comparison between experimental measurements and lattice calculations of
nucleon structure.  The axial charge and momentum fraction of the nucleon
represent the state of affairs with the former observable providing an example of a
potential success of current calculations and the latter an example of
a challenge to future calculations.
\paragraph{$\left<x\right>_{u-d}$}
Extensive quenched calculations of $\left< x \right>_{u-d}$~\cite{Gockeler:2002ek} 
have shown very little dependence on
$m_\pi$ while overestimating the experimental result by nearly a factor of two.  The
first unquenched calculations~\cite{Dolgov:2002zm} confirmed the earlier quenched results and
lead to the suggestion that sizable chiral corrections could accommodate both
lattice calculations and experimental measurements~\cite{Detmold:2001jb}. 
Recent calculations with lighter
quark masses~\cite{Gockeler:2004vx,Ohta:2004mg} have not resolved this discrepancy, however one recent quenched calculation~\cite{Gurtler:2004ac}
shows a significant but unconfirmed shift toward the experimental result.
\paragraph{$g_A$}
Lattice calculations of the nucleon axial coupling are beginning to mature.  
In particular recent calculations with chiral actions allow for a non-perturbative
renormalization of $g_A$.  This observable has been shown to have sizable 
finite size corrections for light quark masses~\cite{Sasaki:2003jh,Khan:2004vw}, 
however current calculations have reached large enough volumes that such effects 
appear under control.  Furthermore simple linear extrapolations, in $m_\pi^2$, of 
the lightest calculations~\cite{Renner:2004ck} give estimates of $1.23(2)$ to $1.26(2)$ (using 3 to 6 of the lightest points) the latter of which agrees with 
the experimental measurement within the statistical errors. However detailed 
study of the chiral behavior is needed to reliably estimate the systematic errors 
in such extrapolations.
\section{Conclusions}
Lattice calculations of nucleon structure are beginning to realize their promise
to elucidate QCD and make contact with the experimental programs.  Recent
calculations are painting a qualitative three dimensional picture 
of nucleon structure revealing a significant $x$ dependence of the transverse
size of the nucleon.  Quantitative calculations of moments of parton 
distributions are progressing, in particular the calculation of $g_A$ may
soon reach a few percent accuracy.
%thus providing strong encouragment to face
%the remaining challenges.
%
\bibliographystyle{aipproc}
\bibliography{dis}
\end{document}